\documentclass{article}
\usepackage[T1]{fontenc}
\usepackage[utf8]{inputenc}
\usepackage{mathpazo}
\usepackage{microtype}
\usepackage{parskip}
\usepackage{amsmath,amssymb,amsthm,mathtools}
\usepackage{graphicx}
\usepackage{xcolor}
\usepackage{float}
\usepackage{tikz}
\usetikzlibrary{positioning,calc,decorations.pathreplacing}
\usepackage{algorithm}
\usepackage{algpseudocode}
\usepackage{booktabs}
\usepackage{caption}
\usepackage{subcaption}
\usepackage{hyperref}
\usepackage{cleveref}
\usepackage[backend=biber,backref=true,style=numeric,giveninits=true,natbib=true,maxbibnames=12]{biblatex}
\newtheorem{theorem}{Theorem}[section]

\newtheorem{remark}[theorem]{Remark}

\DeclareMathOperator{\EE}{\mathbb{E}}

\title{Optimal transport with a density-dependent cost function}

\author{
Zichu Wang\thanks{Courant Institute of Mathematical Sciences, New York University \href{mailto:zw3409@nyu.edu}{\texttt{zw3409@nyu.edu}}}
\and
Esteban G. Tabak\thanks{Courant Institute of Mathematical Sciences, New York University, \href{mailto:tabak@cims.nyu.edu}{\texttt{tabak@cims.nyu.edu}}}
}
\date{}

\begin{document}

\maketitle

\begin{abstract}

A new pairwise cost function is proposed for the optimal transport barycenter problem, adopting the form of the minimal action between two points, with a Lagrangian that takes into account an underlying probability distribution. Under this notion of distance, two points can only be close if there exist paths joining them that do not traverse areas of small probability. A framework is proposed and developed for the numerical solution of the corresponding data-driven optimal transport problem. The procedure parameterizes the paths of minimal action through path dependent Chebyshev polynomials and enforces the agreement between the paths' endpoints and the given source and target distributions through an adversarial penalization. The methodology and its application to clustering and matching problems is illustrated through synthetic examples.

\end{abstract}

\section{Introduction}

The optimal transport problem [OT] seeks a map between two distributions $p_{0,1}$ on a $d$-dimensional manifold $\mathcal{X}$ that minimizes a total transportation cost,
$$ C = \min_T \EE\left[c\left(x, T(x)\right)\right] \quad \hbox{s.t.} \quad T\#p_0 = p_1, $$
where $c$ is an externally provided pairwise transportation cost, and the symbol ``$\#$'' denotes the push-forward measure, such that for all measurable sets $A$,
$$ p_1\left[A\right] = p_0\left[T^{-1}(A)\right]. $$

The following are some of the many uses of OT:
\begin{itemize}
\item The map $T$ pairing two distributions is instrumental for the solution of matching problems \cite{Galichon2016};

\item The total cost $C$ provides a ``horizontal'' notion of distance between $p_{0,1}$, as opposed to ``vertical'' measures of discrepancy based on the pointwise comparison of their values, such as their $L^2$-distance and relative entropy \cite{Santambrogio2015};

\item When one of the distributions is known, the map $T$ estimates the other distribution through the change of variable formula. Similarly, when one of the distributions can be easily sampled, the map $T$ permits simulating the other, acting as a generative model \cite{TabakTrigila2016};

\item Considering $y = T(x)$ as the end point of a path linking $x$ and $y$ allows one to interpolate between the two distributions \cite{McCann1997}; 

\item The optimal transport problem is a critical building block for the more general optimal transport barycenter problem, which permits estimating and simulating conditional distributions \cite{TabakTrigila2018}.

\end{itemize}

The choice of a pairwise cost $c$ is critical in most of these applications, as it determines the optimal map among the infinitely many pushing forward $p_0$ to $p_1$. Thus the distance $C$ between the two distributions, their pairing through $T$ and the interpolation between them will only be meaningful if the pairwise cost $c$ is natural for the problem in hand. Default choices, such as the canonical
\begin{equation}
 c(x, y) = \frac{1}{2}\|y - x\|^2 
 \label{canonical}
\end{equation}
in normed spaces, may fail to capture some field-dependent notion of ``true'' distance between pairs of points. In Riemannian manifolds, for instance, a natural choice for $c$ is the squared geodesic distance between $x$ and $y$. Then the corresponding interpolation between distributions yields geodesics in Wasserstein space \cite{Otto2001}.

In problems in data analysis, it is natural to measure the distance between two points along the path joining them, provided that the areas that these paths traverse in phase space have a relatively large probability. It would not make sense to call two points ``close'' if connecting one to the other requires going through states that are unrealizable or highly unlikely. This was the logic used in \citep{SapienzaGroismanJonckheere2018} to define the \emph{Fermat distance} between points, a data-based construction based on realizable paths.

A natural class of cost functions is the action derivable from a Lagrangian,
\begin{equation}
 c(x, y) = \min_{w(t)} \int_{0}^{1} L(\dot{w},w) \, dt,
\quad w(0) = x, \ w(1) = y. 
\label{action}
\end{equation}
The canonical cost in (\ref{canonical}) is a particular case of this, with Lagrangian $ L = \frac{1}{2} \|\dot{w}\|^2$. We can easily modify this Lagrangian to make the transportation cost depend on the likelihood in phase space, writing
\begin{equation}
  L = \frac{1}{2} \frac{\|\dot{w}\|^2}{\rho(w)^{\alpha}}, \quad \alpha > 0,
  \label{Lagrangian}
\end{equation}
where $\rho$ is a background probability density. We conceptualize $\rho$ as the mixture of the infinitely many distributions among which we may perform pairwise optimal transport. Thus we introduce a family $p_z(w) = p(w\mid z)$ from a joint distribution $\pi(w,z)$ through $p(w\mid z)=\pi(w,z)/\pi(\mathcal{X},z)$, and define
$$ \rho(w) = \int \pi(w, z) \ dz. $$

The cost function in (\ref{action}) with Lagrangian (\ref{Lagrangian}) combines the intuition behind the Fermat distance with the convenience of the well-grounded framework of Lagrangian dynamics. 
This article develops a methodology to implement this cost function in the data-driven optimal transport problem. More generally, the methodology addresses any cost function that adopts the form of an action, which, when specialized to actions that depend on an underlying density as in (\ref{Lagrangian}) yields meaningful applications in data science. The main ingredients of the procedure are the following.

Consider first the problem of finding the transportation cost between two points $x$ and $y$, i.e. solving the minimal action problem in (\ref{action}). In the context of solving the optimal transport problem, finding the value of $c(x, y)$ is not enough: we should be able to differentiate $c$ with respect to $x$ and $y$. Having defined $c$ as an action makes this particularly straightforward, since
$$ \nabla_y c(x, y) = m(1), \quad \nabla_x c(x, y) = -m(0), $$
where the \emph{momentum} $m(t)$ is defined through
$$ m(\tau) = \frac{\partial L}{\partial \dot{w}} \Big|_{t = \tau} $$
In order to have computational access to $m(t)$, we need a closed form expression for $w(t)$, for which we choose to approximate it through Chebyshev polynomials with coefficients $\{a_k\},\ k \in \{1, \ldots, N_{\text{cheb}}\}$. Then the problem reduces to minimizing the action $c(x, y)$ over these coefficients.

Yet we need more than the optimal path between a pair of points: we need a full family of paths connecting the support of $p_0$ and $p_1$. We represent this family by making the $\{a_k\}$ depend on a parameter $s \in \mathcal{S}\subset\mathbb{R}^d$, draw $m$ sample paths parameterized by anchors $\{s^j\}, \ j \in \{1, \ldots, m\}$, and approximate the functions $\{a_k(s)\}$ through reproducing kernels $K:\mathcal{S}\times\mathcal{S}\to\mathbb{R}$.

The paths' initial and final points $\{w(s,0), w(s,1)\}$ are not known a priori: the total transportation cost must be minimized over their location too, subject to the constraints that $w(s,0) \sim p_0$, $w(s,1) \sim p_1$. We consider data-driven scenarios where $p_{0,1}$ are known through samples
$$ X_0=\{x_0^i\}_{i=1}^{n_0}, \qquad X_1=\{x_1^i\}_{i=1}^{n_1}, $$
for which we impose the condition that the distributions underlying the $\{w(s^{j},0)\}$ and $\{x_0^i\}$ are the same, and similarly for $\{w(s^{j},1)\}$ and $\{x_1^i\}$. To enforce these, we penalize the difference between distributions through an instance of an adversarial procedure developed in \cite{LipnickTabakTrigilaWangYeZhao2025} for the optimal transport barycenter problem.

\subsection{Related work}

Our choice of a cost function defined through an optimal path that avoids areas of low probability was inspired by the Fermat distance \citep{SapienzaGroismanJonckheere2018}. Yet the latter is heavily data-based --which is also one of its virtues-- and hard to use as a cost for the optimal transport problem in a computational setting. A minimal action-based cost with density-dependent Lagrangian is much more suitable for this. 

The literature on the OT problem starts with the classical work in Monge and Kantorovich \citep{Monge1781,Kantorovich1942}, separated by over a century, and continues afterward with very many significant contributions, which we cannot possibly summarize here. Most use explicitly defined cost functions, typically the squared Euclidean distance. Action-based costs appear naturally in the application of OT to Riemannian geometry (see \citep{Villani2009,Villani2003} and references therein), but we are not aware of work addressing their numerical implementation or making them depend on an ambient density. We hope that our work can help fill this gap and be of use in a geometrical context. Another connection between optimal transport and minimal action is the classical work in \citep{BenamouBrenier2000}, which used a fluid mechanical interpretation to solve OT problems in a Lagrangian framework.

\subsection{Plan of the article}

After this introduction, section \ref{sec:optimal-path-two-points} addresses the problem of computing the minimal action path and differentiating it with respect to its two end points, using Chebyshev polynomials. Section \ref{sec:ot-two-dists} lifts this construction to the solution of the corresponding data-based optimal transport problem, incorporating two new elements: a reproducing kernel Hilbert space for the representation of the path-dependent Chebyshev coefficients and an adversarial formulation to match the paths' endpoints with the given initial and final distributions. Section \ref{sec:general-L} shows that this solution is not limited to the isotropic Lagrangian at hand but extends to more general metrics, which makes it applicable to a broader class of problems. Section \ref{sec:applications} illustrates the advantages of the proposed methodology in two different applications: the clustering of distributions and  matching problems. Finally, section \ref{sec:conclusion} summarizes the work and suggests further avenues of research.

\section{Optimal path between two points}
\label{sec:optimal-path-two-points}

This section solves a relaxation of the minimal-action problem between two points, developing a complete pipeline from the continuous minimal–action model to a computable discrete solver and explains the geometry and numerics behind each component. Recall that the cost of a path depends not only on its length but also on how probable the regions it traverses are. Regions with larger density $\rho$ are easier to traverse; the parameter $\alpha\ge 0$ controls the strength of this preference.

\medskip
\noindent\textbf{Continuous optimization and geometric intuition.}
Given endpoints $x_0 ,x_1\in\mathcal{X}\subset\mathbb{R}^d$, we minimize the action over paths $w:[0,1]\to\mathcal{X}$ with $w(0)=x_0$ and $w(1)=x_1$:
\begin{equation}
c(x_0, x_1) = \min_{w} \int_0^1 L\big(\dot w(t), w(t)\big) dt.
\label{eq:pair-action}
\end{equation}
We use as Lagrangian the squared speed measured in a position dependent, isotropic metric with weight $\tfrac12\rho(w)^{-\alpha}$,
\begin{equation}\label{eq:lagrangian}
L(\dot w,w) = \tfrac12 \|\dot w\|^2 \rho(w)^{-\alpha}.
\end{equation}
When $\alpha=0$ the optimal path is a straight line, while for $\alpha>0$ the path bends towards regions where $\rho$ is large, since high-density corridors reduce the value of the action.

Anticipating that in our application to optimal transport $x_0$ and $x_1$ will not be known beforehand, rather than enforcing $w(0)=x_0$ and $w(1)=x_1$ exactly, we allow the endpoints to vary and penalize their deviation from $x_0$ and $x_1$: 
\[
\mathcal{J}(a,w_0,w_1) = S[w_{N_{\text{cheb}}}] + \lambda\big(\|w_0-x_0\|^2 + \|w_1-x_1\|^2\big),
\qquad
S[w_{N_{\text{cheb}}}] = \int_0^1 L\big(\dot w_{N_{\text{cheb}}}(t), w_{N_{\text{cheb}}}(t)\big)\,dt,
\]
with $w_0=w_{N_{\text{cheb}}}(0)$ and $w_1=w_{N_{\text{cheb}}}(1)$. The parameter $\lambda>0$ controls a trade-off between enforcing the condition that $w_0 = x, w_1 = y$ and minimizing the action. Our procedure alternates between updates of $(w_0,w_1)$ and of the interior coefficients $a$.

\subsection{Chebyshev parameterization with hard endpoints}
\label{subsec:cheb-param}
Chebyshev bases provide high approximation power and good conditioning for functions defined on the interval $[0,1]$. We adopt endpoint vanishing basis functions, so that hard endpoint conditions hold identically and no projection is needed during optimization. Let $T_k$ be the Chebyshev polynomials of the first kind and set $z=2t-1\in[-1,1]$. Define
\[
\phi_k(t) = T_k(2t-1) - T_k(-1)(1-t) - T_k(1)t = T_k(2t-1) - (-1)^k(1-t) - t
\]
and, for truncation order $N_{\text{cheb}}\ge 2$,
\[
w_{N_{\text{cheb}}}(t) = (1-t)\ w_0 +  t\ w_1 + \sum_{k=2}^{N_{\text{cheb}}} a_k \phi_k(t), \qquad a_k\in\mathbb{R}^d,
\]
so that the conditions $w_{N_{\text{cheb}}}(0) = w_0$ and $w_{N_{\text{cheb}}}(1)= w_1$ hold identically. These definitions lift naturally to $d$ dimensions by adopting vector coefficients $a_k\in\mathbb{R}^d$. Then
\[
\phi_k'(t) = 2k U_{k-1}(2t-1) + (-1)^k - 1,
\]
where the $\{U_m\}$ are Chebyshev polynomials of the second kind with $U_m(1)=m+1$ and $U_m(-1)=(-1)^m(m+1)$. In particular,
\[
\phi_k'(1)=2k^2 + (-1)^k - 1,\qquad
\phi_k'(0)=2k^2(-1)^{k-1} + (-1)^k - 1,
\]
and the path velocity is given by
\[
\dot w_{N_{\text{cheb}}}(t) = (w_1 - w_0) + \sum_{k=2}^{N_{\text{cheb}}} a_k \phi_k'(t).
\]
For efficiency and numerical stability, we precompute and cache $\{\phi_k(t_j),\phi_k'(t_j)\}$ at quadrature nodes $\{t_j\}$.

\subsection{Discrete action, gradients and equations of motion}
\label{subsec:discrete-and-grads}

Let $\{(t_r,q_r)\}_{r=1}^{M}$ be Gauss–Legendre nodes and weights on $[0,1]$, and set
$w_r:=w_{N_{\text{cheb}}}(t_r)$ and $\dot w_r:=\dot w_{N_{\text{cheb}}}(t_r)$. The resulting discrete action is
\[
S_{\mathrm{disc}}(a) = \sum_{r=1}^{M} q_r \tfrac12 \|\dot w_r\|^2 \rho(w_r)^{-\alpha}.
\]
Using
\[
\frac{\partial L}{\partial \dot w}(\dot w,w)=\rho(w)^{-\alpha}\dot w,
\qquad
\nabla_w L(\dot w,w) = -\tfrac{\alpha}{2}\|\dot w\|^2 \rho(w)^{-\alpha} \nabla\log\rho(w),
\]
the gradient over the coefficients $a$ is given by
\[
\nabla_{a_k} S_{\mathrm{disc}}
=
\sum_{r=1}^{M} q_r\Big[
\rho(w_r)^{-\alpha}\dot w_r \phi_k'(t_r)
-\tfrac{\alpha}{2}\|\dot w_r\|^2 \rho(w_r)^{-\alpha}\nabla\log\rho(w_r) \phi_k(t_r)
\Big],
\quad k=2,\dots,N_{\text{cheb}}.
\]

\paragraph{Endpoint gradients via calculus of variations.}
For $S[w_{N_{\text{cheb}}}]=\int_0^1 L(\dot w_{N_{\text{cheb}}}(t),w_{N_{\text{cheb}}}(t)) dt$ and a variation with
$\delta w(0)=\delta w_0$, $\delta w(1)=\delta w_1$,
\[
\delta S
=\Big\langle \frac{\partial L}{\partial \dot w},\delta w\Big\rangle\Big|_{t=1}
-\Big\langle \frac{\partial L}{\partial \dot w},\delta w\Big\rangle\Big|_{t=0}
+\int_0^1 \Big\langle -\frac{d}{dt}\frac{\partial L}{\partial \dot w}
+\frac{\partial L}{\partial w},\delta w\Big\rangle dt.
\]
At a stationary path the integral term vanishes, hence
\[
\nabla_{w_1} S = \nabla_{\dot{w}} L \Big|_{t=1},
\qquad
\nabla_{w_0} S =- \nabla_{\dot{w}} L \Big|_{t=0}.
\]
With the expressions above,
\[
\nabla_{w_1} c(w_0, w_1)= \nabla_{w_1} S 
=\rho(w_1)^{-\alpha}\dot w_{N_{\text{cheb}}}(1),
\qquad
\nabla_{w_0} c(w_0, w_1)= \nabla_{w_0} S
=-\rho(w_0)^{-\alpha}\dot w_{N_{\text{cheb}}}(0),
\]
where
\[
\dot w_{N_{\text{cheb}}}(1)=(w_1- w_0)+\sum_{k=2}^{N_{\text{cheb}}} a_k \phi_k'(1),
\qquad
\dot w_{N_{\text{cheb}}}(0)=(w_1- w_0)+\sum_{k=2}^{N_{\text{cheb}}} a_k \phi_k'(0).
\]

\paragraph{Objective with endpoint penalty.}
For $\mathcal{J}=S+\lambda(\|w_1-x_1\|^2+\|w_0-x_0\|^2)$ with $w_0=w_{N_{\text{cheb}}}(0)$ and $w_1=w_{N_{\text{cheb}}}(1)$,
\[
\nabla_{w_1} \mathcal{J}
=\rho(w_1)^{-\alpha}\dot w_{N_{\text{cheb}}}(1)+2\lambda(w_1-x_1),
\qquad
\nabla_{w_0} \mathcal{J}
=-\rho(w_0)^{-\alpha}\dot w_{N_{\text{cheb}}}(0)+2\lambda(w_0-x_0).
\]
We perform gradient descent with step size $\eta>0$,
\[
w_0 \leftarrow w_0 - \eta\,\nabla_{w_0} \mathcal{J},\qquad
w_1 \leftarrow w_1 - \eta\,\nabla_{w_1} \mathcal{J},\qquad
a_k \leftarrow a_k - \eta\,\nabla_{a_k} S_{\mathrm{disc}}\ (k=2,\dots,N_{\text{cheb}}).
\]

\subsection{An example with Gaussian background distribution}
\label{subsec:gaussian-example}

When $\rho$ is a standard Gaussian on $\mathbb{R}^d$, we have $\rho(w)\propto \exp(-\tfrac12\|w\|^2)$ and $\nabla\log\rho(w)=-w$. Then, adopting $\alpha = 1$ yields
\[
\nabla_{a_k} S_{\mathrm{disc}}
=
\sum_{r=1}^{M} q_r\Big[
\rho(w_r)^{-1}\dot w_r \phi_k'(t_r)
+\tfrac12\|\dot w_r\|^2 \rho(w_r)^{-1} w_r \phi_k(t_r)
\Big],\quad k=2,\dots,N_{\text{cheb}},
\]
\[
\nabla_{w_1} \mathcal{J}
=\rho(w_1)^{-1}\Big((w_1-w_0)+\sum_{k=2}^{N_{\text{cheb}}} a_k \phi_k'(1)\Big)+2\lambda(w_1-x_1),
\]
\[
\nabla_{w_0} \mathcal{J}
=-\rho(w_0)^{-1}\Big((w_1-w_0)+\sum_{k=2}^{N_{\text{cheb}}} a_k \phi_k'(0)\Big)+2\lambda(w_0-x_0).
\]
Since path segments near the origin have smaller action, the optimal path bends toward the origin, as can be seen in the examples of figure ~\ref{fig:single-point}.

In order to validate the procedure, we compare its results with the numerical solution of the Euler–Lagrange equations:
\[
\frac{d}{dt}\Big(\nabla_{\dot{w}} L \Big)-\nabla_w L = 0 .
\]
For $L=\tfrac12\|\dot w\|^2\rho(w)^{-1}$, one obtains
\[
\nabla_{\dot{w}} L = \rho(w)^{-1}\dot w,
\qquad
\nabla_{w} L = \tfrac12\|\dot w\|^2\rho(w)^{-1}w,
\qquad
\frac{d}{dt}\rho(w)^{-1}=\rho(w)^{-1}(w\cdot \dot w),
\]
leading to
\[
\ddot w+(w\cdot \dot w)\dot w-\tfrac12\|\dot w\|^2 w=0 .
\]
We solve the Euler–Lagrange two–point BVP by single shooting:
\begin{enumerate}
\item Initialize the unknown initial velocity with \(v^{(0)}=x_1-x_0\).
\item For a given \(v\), integrate the Euler-Lagrange ODE on \([0,1]\) using an adaptive Runge–Kutta scheme to obtain \(w(1;x_0,v)\).
\item Minimize the terminal mismatch \(J(v)=\tfrac12\|w(1;x_0,v)-x_1\|^2\) via a derivative-free line search with backtracking until \(J(v)\le\varepsilon\) (\(\varepsilon=10^{-8}\) here).
\item With \(v^\star\), reintegrate to produce the path \(w(t)\) that hits \(x_1\) at \(t=1\).
\end{enumerate}
As can be seen in the plot, the two procedures produce virtually identical results.

\begin{center}
\includegraphics[scale=0.5]{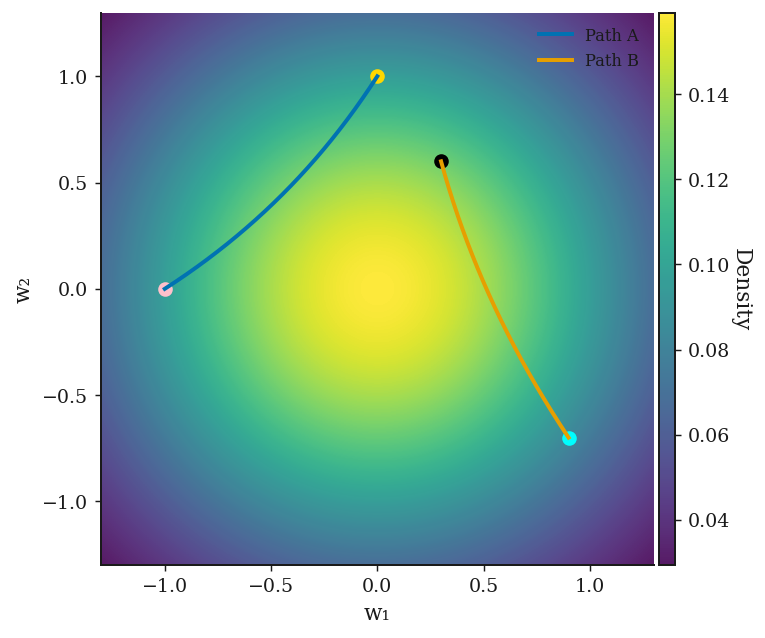}
\captionof{figure}{Two single-pair paths under a Gaussian background. Chebyshev order $N_{\text{cheb}}=10$, Gauss–Legendre nodes, $M=50$, endpoint penalty $\lambda=10^{5}$. The trajectory obtained by solving the Euler–Lagrange boundary-value problem coincides with the trajectory produced by our Chebyshev-based discretization and gradient solver; the two corresponding curves overlap in the plot.}
\label{fig:single-point}
\end{center}

\section{Optimal transport between two distributions}
\label{sec:ot-two-dists}
We now lift the pairwise construction of Section~\ref{sec:optimal-path-two-points} to the optimal transport between distributions. At the population level we seek a family of paths
\[
w(s,\cdot):[0,1]\to\mathcal{X},\qquad s\in\mathcal{S},
\]
that minimizes the average action while pushing $p_0$ to $p_1$:
\[
\min_{w}\ \int_0^1 \int_{\mathcal{S}} L\big(\dot w(s,t), w(s,t)\big) d\pi(s) dt
\quad\text{s.t.}\quad
w(\cdot,0) \# \pi = p_0,\qquad w(\cdot,1) \# \pi = p_1,
\]
where $\pi(s)$ is the --arbitrary-- distribution from which we sample the $\{s^i\}$. 
Equivalently, writing $p_t$ for the law of $w(s,t)$ and $v_t$ for its velocity field, one has the continuity equation 
\[
\partial_t p_t + \nabla\cdot(p_t \ v_t)=0 \quad v_t(\cdot)=\dot w(s,t)
\]
and boundary conditions $p_0, p_1$. Solving this exactly is intractable, so we adopt a data-driven relaxation that preserves the action structure and the endpoint gradient formulas developed earlier.
Given samples $X_0=\{x_0^i\}_{i=1}^{n_0}$ and $X_1=\{x_1^i\}_{i=1}^{n_1}$, we relax the constraints by penalization and optimize over $m$ sample paths $\{w^{j}(\cdot)\}$, each attached to a sample $s^j$:
\[
\min_{\{w^{j}(\cdot)\}}
\ \sum_{j=1}^{m} \int_{0}^{1} L\big(w^{j}(t), \dot w^{j}(t)\big)\ dt
\ +\ \lambda\ \left[R\big(\{w_0^{j}\}, \{x_0^{i}\}\big) + R\big(\{w_1^{j}\}, \{x_1^{i}\}\big) \right],
\]
where the penalty function $R$ enforces alignment in distribution of the endpoint clouds with the empirical source/target sets.
We determine $R$ through a methodology developed in \cite{LipnickTabakTrigilaWangYeZhao2025}, which we summarize here for the $\{w(s^j,1)\}$ and $\{x_1^i\}$, since it applies without changes to the paths' other end. We will assume for simplicity that $n_0 = n_1 = m$, i.e. that the number of samples available for the initial and final distributions and  for the sample paths are the same (This is not required, since the samples must agree in distribution, not point-wise.)  

We introduce a pair of variables $\{y, z\}$, where $y$ is either $w_1$ or $x_1$ and $z$ is a binary variable that specifies which of the two holds. Thus we have $m$ sample pairs of the form $(w_1^j, 0)$ and other $m$ of the form $(x_1^i, 1)$. Then we re-formulate the equality in distribution of $\{w_1^j\}$ and $\{x_1^i\}$ as the requirement that $y$ be independent of $z$. The latter condition is itself equivalent to requiring that all measurable functions $g(y)$ and $f(z)$ be uncorrelated. The empirical version of this states that, for any two functions $g$ and $f$,
$$ \sum_i f(z^i) = 0 \ \Rightarrow \ \sum_i g(y^i) f(z^i) = 0 .$$
Now, for a binary variable $z$, there exist only one function $f$ with zero mean, except for the irrelevant choice of a sign and an amplitude; in our case with $n_1 = m$, this function is
\[
f(z)=
\begin{cases}
\frac{1}{m}, & z=0\\
-\frac{1}{m}, & z=1 .
\end{cases}
\]
We restrict $g(y)$ to the space spanned by a finite set of functions $G^l(y)$, which can be chosen as rich as needed, for instance through reproducing kernel Hilbert spaces. Since the examples we use in this article to demonstrate the procedure involve only Gaussian distributions, fully described by their mean vectors and covariance matrices, it suffices to consider linear and quadratic functions, which in our two dimensional examples yield the five independent functions, $\{G^l(y)\}=\{y_1, y_2, y_1^2, y_1 y_2, y_2^2\}$, which we assemble into the matrix $G\in\mathbb{R}^{n\times 5}$, with $G_{ij} = G^j(y^i)$.

We will penalize $(\sum_i g(y^i) f(z^i))^2$, so it is important that this be minimized based on the correlation between $g$ and $f$, not on the magnitude of $g$. To this end, we remove the mean of each column of $G(y)$, replace $G(y)$ by an orthogonal matrix $Q_1(y)$ spanning the same space, and write
$$ g(y) = Q_1(y) b, \quad \|b\| = 1, $$
which guarantees that $\|g\| = 1$. The reason for the index ``$1$'' in $Q_1$ is that the orthogonalization of $G$ is based on the empirical inner product defined over the $y_1 = \{w_1^i\} \cup \{x_1^j\}$ (A similar $Q_0(y)$ is used at the other end.)

Then we define
$$ R\big(\{w_1^{j}\}, \{x_1^{i}\}\big) = \max_{\|b\| = 1} \|f^\top Q_1(y_1) b\|^2 = \|f^\top Q_1(y_1)\|^2 .$$

\subsection{Cost Part, RKHS parameterization and discretization}
\label{subsec:rkhs-parameterization}

\paragraph{RKHS parameterization} To couple information across many paths in a statistically efficient manner, we place the time-varying coefficients of the Chebyshev expansion in a reproducing kernel Hilbert space (RKHS) over an index set $\mathcal{S}$. Let $K:\mathcal{S}\times\mathcal{S}\to\mathbb{R}$ be a positive definite kernel with associated RKHS $\mathcal{H}$. For each mode $k\ge 2$ we model the coefficient field $a_k:\mathcal{S}\to\mathbb{R}^d$ in the finite span of kernel sections at anchors $\{s^{j}\}_{j=1}^{m}\subset\mathcal{S}$:
\[
a_k(s) = \sum_{j=1}^{m} \theta_{k,j} K(s,s^{j}),\qquad k=2,\dots,N_{\text{cheb}},
\]
with parameters $\theta_{k,j}\in\mathbb{R}^d$. This parameterization shares statistical strength across anchors and approximates smooth coefficient fields in $\mathcal{H}$.

\smallskip \noindent For a fixed anchor $s^{j}$ the path $w^{j}(t):=w(s^{j},t)$ is represented as before by the endpoint-satisfying Chebyshev expansion
\[
w^{j}(t) = (1-t)\ w^{j}(0) + t\ w^{j}(1) + \sum_{k=2}^{N_{\text{cheb}}} a_k(s^{j}) \phi_k(t).
\]

\paragraph{Gradients with respect to coefficients and RKHS parameters}
Let $\{(t_r,q_r)\}_{r=1}^{M}$ be Gauss–Legendre nodes and weights on $[0,1]$. For each anchor $s^{j}$, set $w^{j}_r=w^{j}(t_r)$ and $\dot w^{j}_r=\dot w^{j}(t_r)$. The discrete cost is
\[
\mathcal{S}_{\mathrm{dist}}(\Theta,\{w^{j}(0),w^{j}(1)\})
= \sum_{j=1}^{m}\sum_{r=1}^{M} q_r L\big(\dot w^{j}_r,w^{j}_r\big),
\qquad
L(\dot w,w)=\tfrac12\|\dot w\|^2 \rho(w)^{-\alpha}.
\]
Under the Chebyshev representation above,
the partial derivatives of the path (Jacobian matrices in $\mathbb{R}^{d\times d}$) are
\[
\frac{\partial w^{j}_r}{\partial a_k(s^{j})}=\phi_k(t_r) I_d,
\qquad
\frac{\partial \dot w^{j}_r}{\partial a_k(s^{j})}=\phi_k'(t_r) I_d.
\]
At each node we have,
\[
\frac{\partial L}{\partial \dot w}(\dot w^{j}_r,w^{j}_r)=\rho(w^{j}_r)^{-\alpha} \dot w^{j}_r,
\qquad
\nabla_w L(\dot w^{j}_r,w^{j}_r)=-\tfrac{\alpha}{2}\|\dot w^{j}_r\|^2 \rho(w^{j}_r)^{-\alpha} \nabla\log\rho(w^{j}_r),
\]
so by the chain rule the gradient with respect to the local coefficient field $a_k(s^{j})\in\mathbb{R}^d$ is
\[
\frac{\partial \mathcal{S}_{\mathrm{dist}}}{\partial a_k(s^{j})}
=
\sum_{r=1}^{M} q_r\Big[
\rho(w^{j}_r)^{-\alpha} \dot w^{j}_r \phi_k'(t_r)
-\tfrac{\alpha}{2} \|\dot w^{j}_r\|^{2} \rho(w^{j}_r)^{-\alpha} \nabla\log\rho(w^{j}_r) \phi_k(t_r)
\Big]\in\mathbb{R}^d,
\]
and, componentwise for $\ell=1,\dots,d$,
\[
\frac{\partial \mathcal{S}_{\mathrm{dist}}}{\partial (a_k)_\ell(s^{j})}
=
\sum_{r=1}^{M} q_r\left[
\rho(w^{j}_r)^{-\alpha} (\dot w^{j}_r)_\ell \phi_k'(t_r)
-\frac{\alpha}{2} \|\dot w^{j}_r\|^2 \rho(w^{j}_r)^{-\alpha} (\nabla\log\rho(w^{j}_r))_\ell \phi_k(t_r)
\right].
\]
Since the coefficient fields are shared across anchors via the RKHS expansion
\[
a_k(s)=\sum_{j=1}^{m}\theta_{k,j} K(s,s^{j}),
\]
then
\[
\frac{\partial \mathcal{S}_{\mathrm{dist}}}{\partial \theta_{k,j}}
=
\sum_{j'=1}^{m} K(s^{j'},s^{j})
\frac{\partial \mathcal{S}_{\mathrm{dist}}}{\partial a_k(s^{j'})},
\qquad j=1,\dots,m,\ k=2,\dots,N_{\text{cheb}},
\]
and a first–order update with step size $\eta>0$ is
\[
\theta_{k,j}\leftarrow \theta_{k,j}-\eta \frac{\partial \mathcal{S}_{\mathrm{dist}}}{\partial \theta_{k,j}}.
\]

After including the penalty terms for agreement with the initial and final distributions, the full objective function becomes

%
\[
\mathcal{J}(\Theta,W_0,W_1)
=\sum_{j=1}^{m}\sum_{r=1}^{M} q_r\, L\!\big(\dot w^{\,j}(t_r),w^{\,j}(t_r)\big)
\;+\;\lambda_0\,\bigl\|f^\top Q_{w(0)}\bigr\|^2
\;+\;\lambda_1\,\bigl\|f^\top Q_{w(1)}\bigr\|^2,
\]
where $w^{\,j}$ is the endpoint-satisfying Chebyshev path for anchor $s^j$ and $\Theta$ are the RKHS coefficients for the interior modes.

%
Writing $m(t):=\partial L/\partial \dot w\big(\dot w(t),w(t)\big)$ for the boundary momentum, we have,
for $L=\tfrac12\|\dot w\|^2/\rho(w)$, that
$m(t)=\dot w(t)/\rho\big(w(t)\big)$, and the endpoint gradients are given by
\[
\frac{\partial \mathcal{J}}{\partial w_i(0)}
= -\,\frac{\dot w_i(0)}{\rho\bigl(w_i(0)\bigr)}
+ 2\lambda_0\,\bigl(f^\top Q_{w(0)}\bigr)^\top
\left[f^\top\,\frac{1}{\|\widetilde G\|_F}\,\frac{\partial G}{\partial w_i(0)}\,B_{k_{\mathrm{svd}}}\right],
\]
\[
\frac{\partial \mathcal{J}}{\partial w_i(1)}
= \frac{\dot w_i(1)}{\rho\bigl(w_i(1)\bigr)}
+ 2\lambda_1\,\bigl(f^\top Q_{w(1)}\bigr)^\top
\left[f^\top\,\frac{1}{\|\widetilde G_1\|_F}\,\frac{\partial G_1}{\partial w_i(1)}\,B_{1,k_{\mathrm{svd}}}\right].
\]
(Here $G,G_1,\widetilde G,\widetilde G_1,B_{k_{\mathrm{svd}}},B_{1,k_{\mathrm{svd}}}$ and the feature Jacobians are defined in the Appendix, which also specifies the fixed centering/normalization convention that we use.)

\smallskip
Since the interior coefficients are unaffected by the endpoint penalties, their gradients coincide with the cost-part gradients computed above:
\[
\frac{\partial \mathcal{J}}{\partial a_k(s^{j})}
=\frac{\partial \mathcal{S}_{\mathrm{dist}}}{\partial a_k(s^{j})},
\qquad
\frac{\partial \mathcal{J}}{\partial \theta_{k,j}}
=\sum_{j'=1}^{m} K(s^{j'},s^{j})\,
\frac{\partial \mathcal{S}_{\mathrm{dist}}}{\partial a_k(s^{j'})}.
\]

\paragraph{Updates.}
With step size $\eta>0$,
\[
w_i(0)\leftarrow w_i(0)-\eta\,\frac{\partial \mathcal{J}}{\partial w_i(0)},\qquad
w_i(1)\leftarrow w_i(1)-\eta\,\frac{\partial \mathcal{J}}{\partial w_i(1)},\qquad
\theta_{k,j}\leftarrow \theta_{k,j}-\eta\,\frac{\partial \mathcal{J}}{\partial \theta_{k,j}}.
\]

\subsection{An example}
\label{subsec:example}
\begin{figure}[H]
  \centering
  \includegraphics[width=0.5\linewidth]{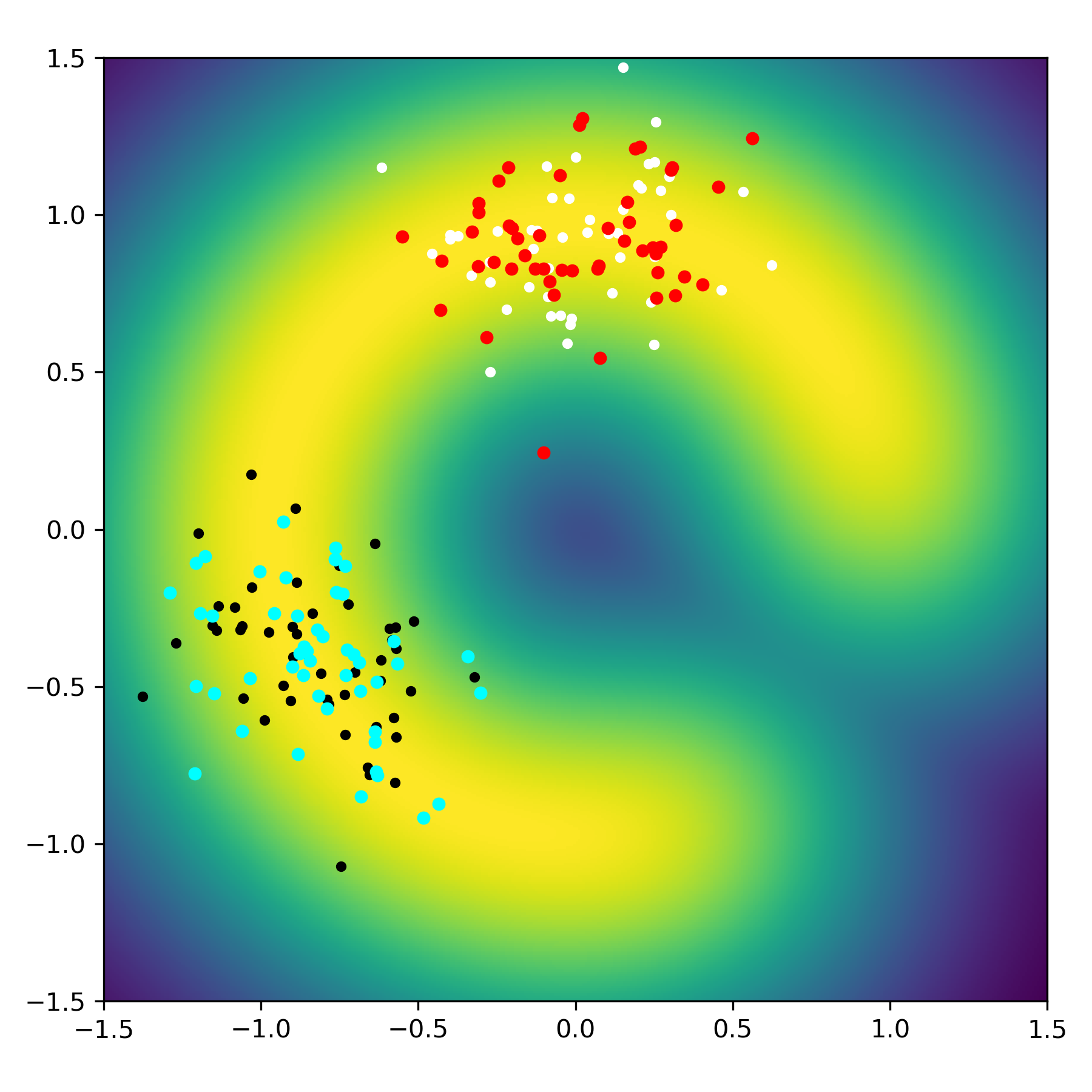}
  \caption{Background shows the constructed density (displayed as $\rho^{1/5}$). The target point clouds are plotted in black and white, and the optimized endpoints are shown in blue and red. Chebyshev order $N_{\text{cheb}}=10$, Gauss–Legendre nodes $M=50$.}
  \label{fig:density-target-optimized}
\end{figure}

\begin{figure}[H]
  \centering
  \begin{subfigure}{0.32\linewidth}
    \includegraphics[width=\linewidth]{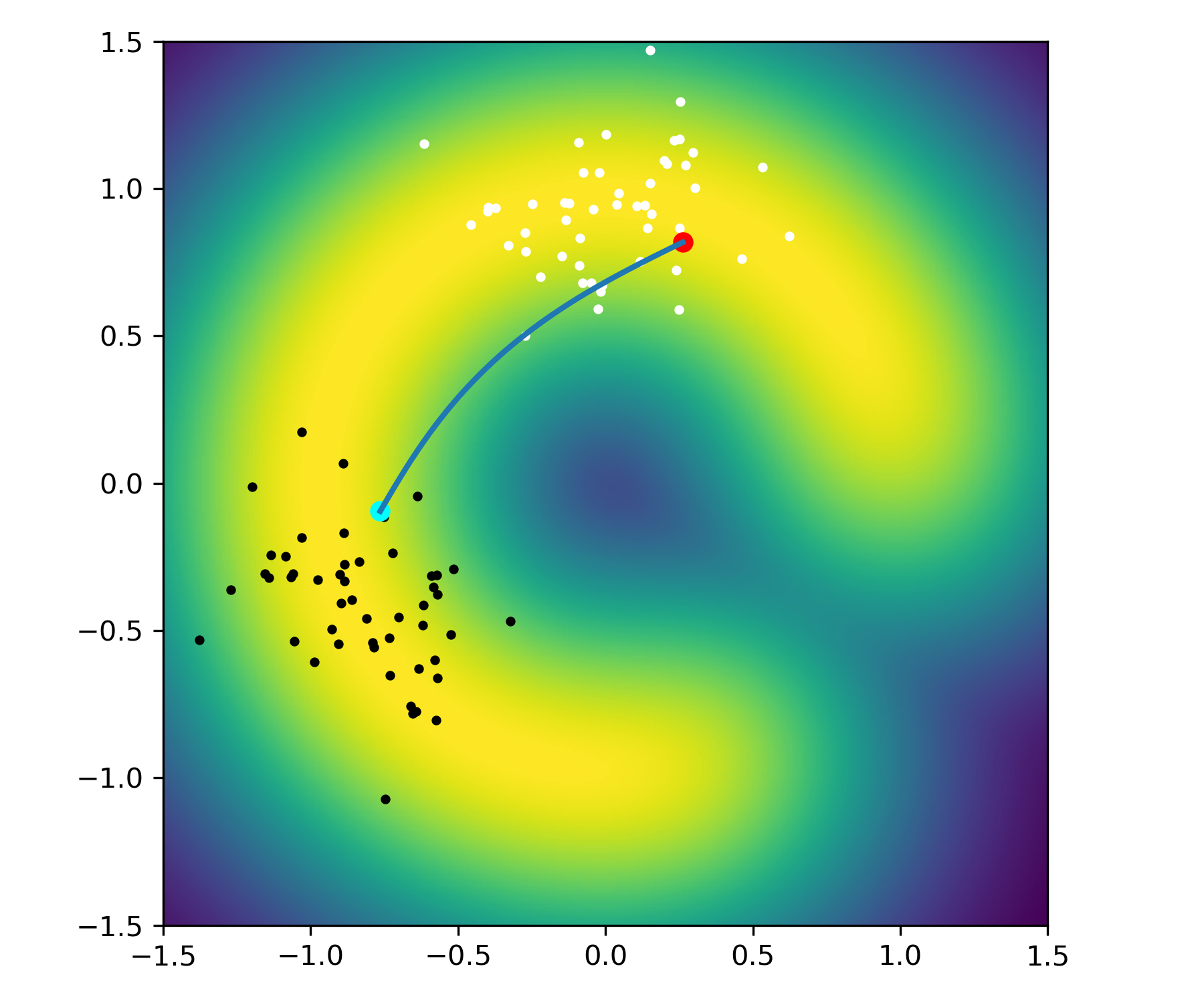}
    \caption{}
    \label{fig:path-009}
  \end{subfigure}\hfill
  \begin{subfigure}{0.32\linewidth}
    \includegraphics[width=\linewidth]{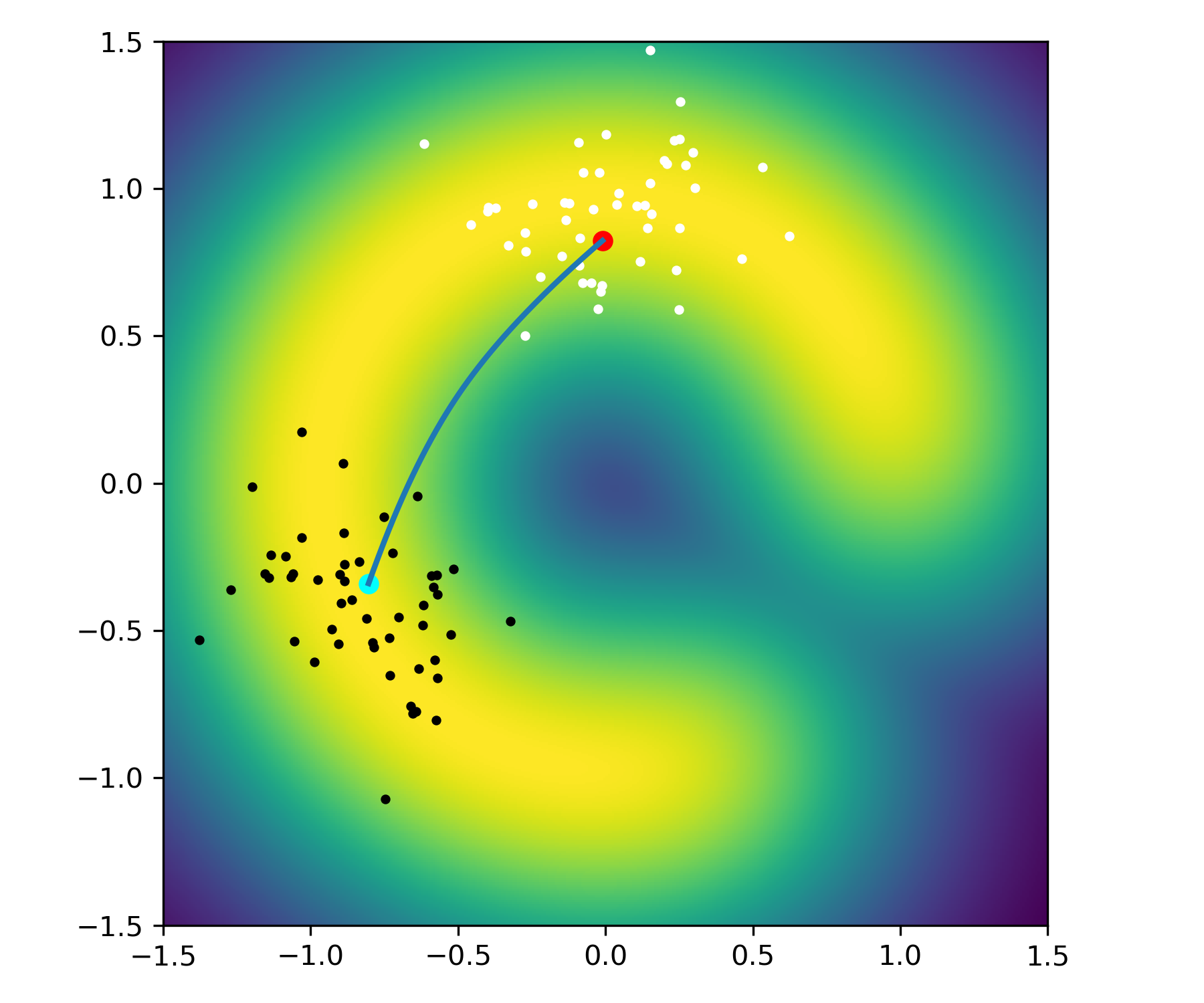}
    \caption{}
    \label{fig:path-013}
  \end{subfigure}\hfill
  \begin{subfigure}{0.32\linewidth}
    \includegraphics[width=\linewidth]{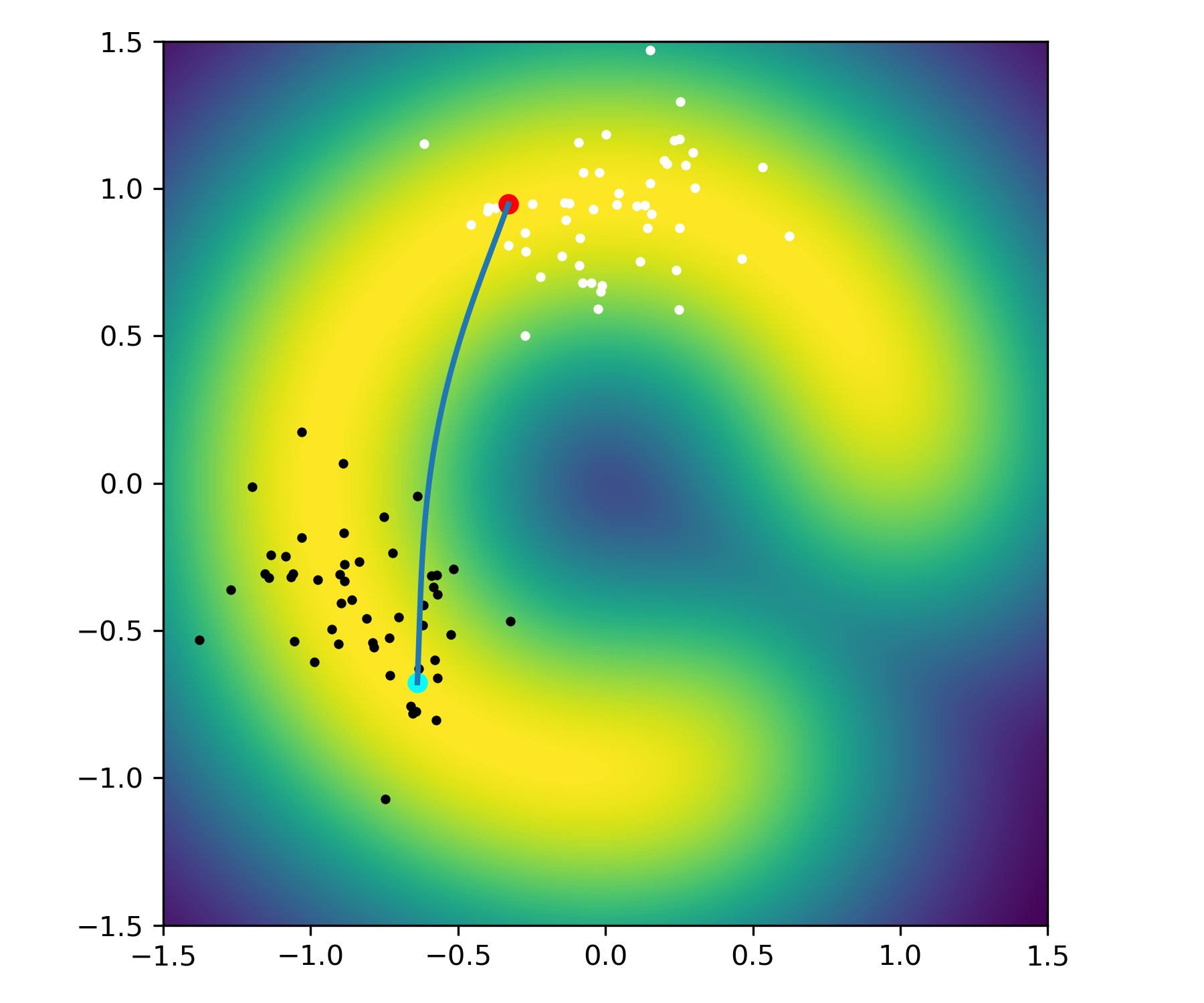}
    \caption{}
    \label{fig:path-045}
  \end{subfigure}
  \caption{Three examples of density–weighted optimal paths; the background shows the constructed density $\rho^{1/5}$.}
  \label{fig:paths-examples}
\end{figure}

We approximate a ring–shaped background density as a continuous mixture of isotropic Gaussians whose centers trace the unit circle,
$c(z)=(\cos z,\sin z)$ for $z\in[Z_{\min},Z_{\max}]$.
Using Gauss–Legendre nodes $\{z_j\}_{j=1}^{N_Z}$ and weights $\{w_j\}_{j=1}^{N_Z}$ mapped to this interval, we set
\[
\rho(w)\approx C_\rho\sum_{j=1}^{N_Z} w_j\,
\exp\!\Big(-\tfrac{\|w-c(z_j)\|^2}{2\sigma^2}\Big),
\qquad
\nabla\rho(w)\approx\frac{C_\rho}{\sigma^2}\sum_{j=1}^{N_Z} w_j\,
\big(c(z_j)-w\big)\,\exp\!\Big(-\tfrac{\|w-c(z_j)\|^2}{2\sigma^2}\Big).
\]
Figure \ref{fig:density-target-optimized} displays the background density (shown through $\rho^{1/5}$ to enhance contrast in low–density regions), the sample points $\{x_{0,1}\}$ provided and the resulting $\{w_{0,1}\}$, while Fig ~\ref{fig:paths-examples} displays three sample optimal paths connecting pairs $\{w_{0,1}^i\}$.

\section{A More General Lagrangian}
\label{sec:general-L}
Even though we specialized our discussion to Lagrangians of the form \eqref{eq:lagrangian}, the methodology proposed extends almost without change to more general actions. In particular, since our Lagrangian was quadratic in $\dot{w}$, we can think of it as deriving from a metric, which can be extended to a more general, not necessarily isotropic metric-based action:
\[
L(\dot w,w)=\|\dot w\|_{G(w)}^{2},\qquad
\|\dot w\|_{G(w)}^{2}:=\dot w^{\top}G(w)\,\dot w,
\]
where $G:\mathcal{X}\to\mathbb{R}^{d\times d}$ is symmetric and positive definite for every $w$. Then the optimal paths are geodesics of the Riemannian metric $G(w)$, which specializes when $G\equiv \tfrac12\,\rho(w)^{-\alpha} I_d$ to our earlier construction, favoring high density corridors.

\paragraph{Discretization and gradients in practice.}
We reuse the Chebyshev parameterization $w_{N_{\text{cheb}}}$ and Gauss–Legendre quadrature introduced earlier. No changes are needed in how the path is represented. Only the local derivatives of $L$ differ:
\[
\frac{\partial L}{\partial \dot w}(\dot w,w)=2\,G(w)\,\dot w,
\qquad
\nabla_w L(\dot w,w) = \mathbf{g}(w;\dot w),
\]
where $\mathbf{g}(w;\dot w)\in\mathbb{R}^d$ collects the metric derivatives along $\dot w$, with components
\[
\bigl[\mathbf{g}(w;\dot w)\bigr]_{\ell}=\dot w^{\top}\Bigl(\frac{\partial G(w)}{\partial w_\ell}\Bigr)\dot w
=\sum_{a,b=1}^{d}\dot w_a\,\dot w_b\,\frac{\partial G_{ab}(w)}{\partial w_\ell}.
\]
These expressions drop seamlessly into the discrete action and yield the coefficient gradients given by the chain rule, while the endpoint gradients of the pairwise cost keep the same simple form:
\[
\nabla_y c(x,y)=m(1),\qquad \nabla_x c(x,y)=-m(0), \qquad m(t) = G\bigl(w(t)\bigr)\,\dot w(t),
\]
so the procedure extends seamlessly to costs derived from general Riemannian metrics.

\section{Two applications}
\label{sec:applications}

Applications of our procedure come in two main flavors. One uses the transportation cost $C(p_0, p_1)$ as a measure of dissimilarity between distributions. Such measures have wide applicability across data science; our illustration here uses them as a tool for grouping sets of distributions into clusters that are consistent with an ambient density $\rho_0$. The other flavor uses not the cost $C$ but the optimal map $y = T(x)$ itself; we illustrate this adopting $T$ as a tool for matching pairs of distributions, a task with applications in many fields, particularly in economics \cite{Galichon2016}.

\subsection{Clustering distributions}
Given several point clouds $\{X^{(g)}\}_{g=1}^{G}$ in $\mathbb{R}^2$, we treat each $X^{(g)}$ as an empirical distribution and define a pairwise dissimilarity using the density–weighted action from Section~\ref{sec:ot-two-dists}. For each pair, we use $k$ independent samples from each distribution as endpoint seeds $W^{(g)}_0,W^{(h)}_1$, and optimize the Chebyshev–RKHS path family to minimize the discrete action under the background density~$\rho$:
\[
D_{g\to h}\;\coloneqq\;\min_{\Theta,\,W^{(g)}_0,\,W^{(h)}_1}
\sum_{r=1}^{M} q_r\,\tfrac12\bigl\|\dot w_{g\to h}(t_r)\bigr\|^2\,\rho\!\bigl(w_{g\to h}(t_r)\bigr)^{-\alpha}
\;+\;\lambda_0\bigl\|f^\top Q_{w(0)}\bigr\|^2
\;+\;\lambda_1\bigl\|f^\top Q_{w(1)}\bigr\|^2,
\]
with $w_{g\to h}$ given by the endpoint-satisfying Chebyshev expansion and coefficients $a_k(s)$ in the finite RKHS span at the anchors. The value at convergence (we use $M=50$ quadrature nodes and $N_{\text{cheb}}=10$) serves as the pairwise cost; we assemble the matrix $D=[D_{g\to h}]$ and run agglomerative hierarchical clustering (average linkage). Because the action penalizes paths that must cross low-density gaps, sets linked by high-density corridors cluster together, while cross-branch sets separate, addressing the failure of Wasserstein distance based couplings in this geometry. 

\begin{figure}[H]
  \centering
  \begin{subfigure}{0.49\linewidth}
    \includegraphics[width=\linewidth]{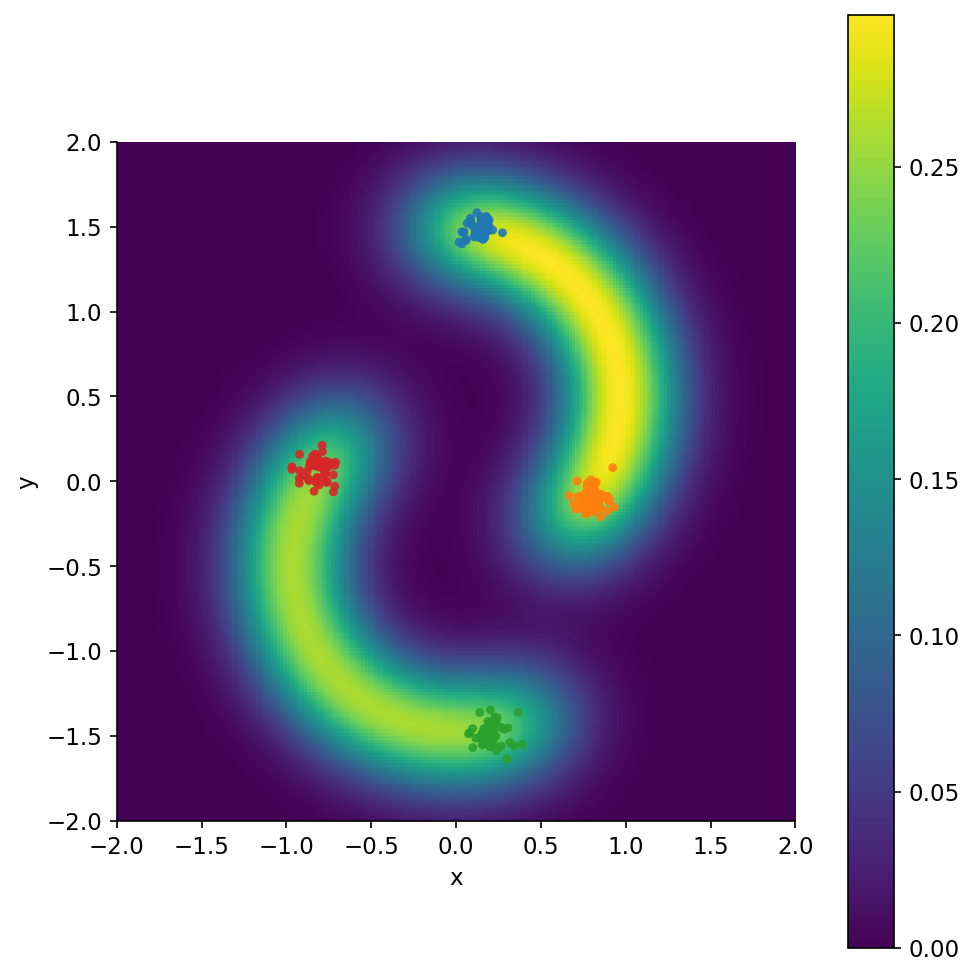}
    \caption{Four point clouds to group and background density.}
    \label{fig:clustering-ours}
  \end{subfigure}
  \begin{subfigure}{0.49\linewidth}
    \includegraphics[width=\linewidth]{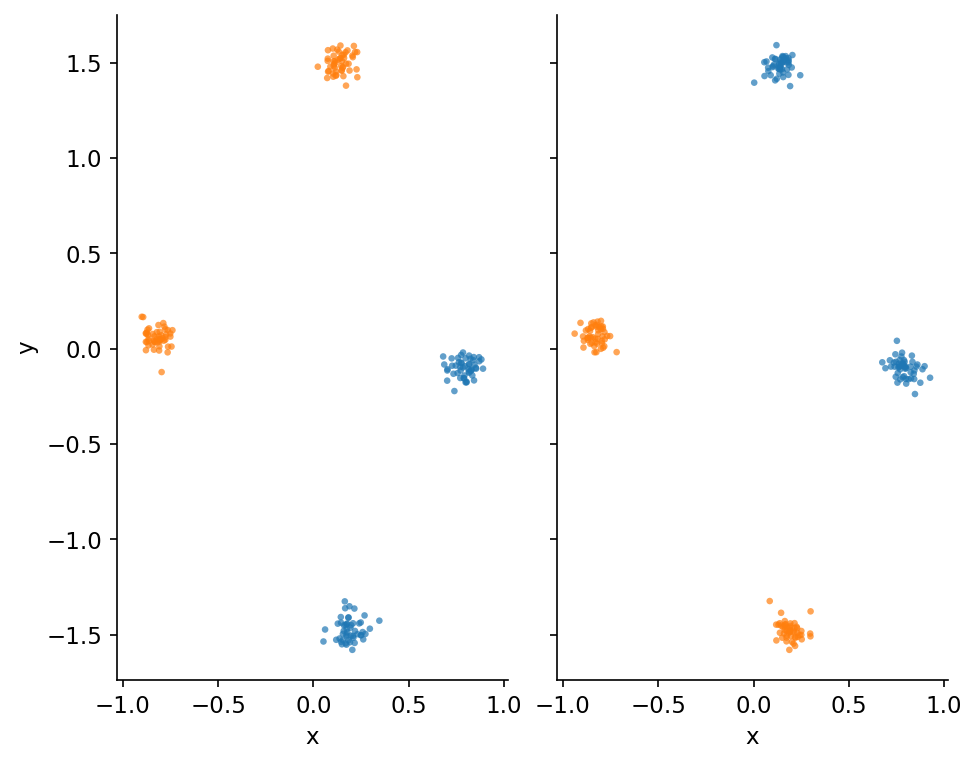}
    \caption{Left panel: grouping under the regular Wasserstein distance. Right panel: corrected grouping under the action-based cost with density-weighted Lagrangian.}
    \label{fig:clustering-w2}
  \end{subfigure}
  \caption{Comparison of pairings: in this example, the new method recovers the intended correspondences, while the Wasserstein endpoint coupling fails to pair the points correctly.}
\label{fig:clustering}
\end{figure}

Fig.~\ref{fig:clustering} illustrates this procedure through the results of applying it to a simple example, with $G = 4$ distributions $X^g$ that we seek to group into two clusters, in a bimodal ambient distribution $\rho_0$ that sorts their pairwise dissimilarity differently from the regular Wasserstein metric: distributions that are close under the Euclidean metric are quite far when the ambient density is taken into account.

\subsection{Matching points}
\begin{figure}[!t]
  \centering
  \begin{minipage}[t]{0.49\linewidth}
    \centering
    \includegraphics[width=\linewidth]{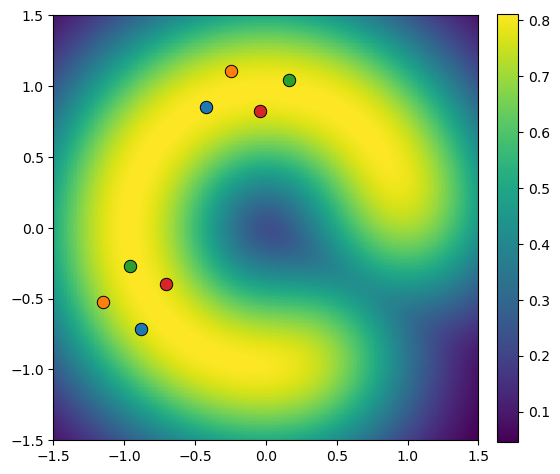}
    \par\smallskip\small (a) New method.
  \end{minipage}\hfill
  \begin{minipage}[t]{0.49\linewidth}
    \centering
    \includegraphics[width=\linewidth]{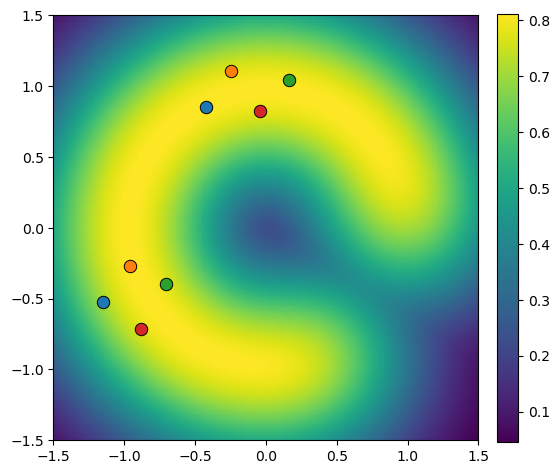}
    \par\smallskip\small (b) Wasserstein coupling.
  \end{minipage}
\vspace{2pt}
  \caption{Comparison of pairings: in this example, the proposed methodology approximates the intended matching between points, corresponding to a rigid rotation, while the Wasserstein endpoint coupling does  that intuition.}
  \label{fig:pairing-comparison}
\end{figure}

We next perform a pairing experiment. We use the same background density as in \ref{subsec:example}, a continuous Gaussian mixture along the unit circle that is large on a thin annulus and exponentially small in the disk's interior. We perform optimal transport between two densities $p_{0,1}$ lying quite far along the disk (see figure \ref{fig:pairing-comparison}), and use the optimal map $T$ as a pairing tool, so that point $x$ is paired to $y = T(x)$. In order to assess the performance of this pairing algorithm, we need to establish a ``ground truth'' to compare it with.
We argue that the most natural interpretation of our $z$-dependent distributions $p(w|z)$ is as resulting from the rigid rotation of a single Gaussian distribution along the ring. Consequently, we should expect the pairing between two distributions with angles $z_{0,1}$ to consist of pairs $\{x_0^i, x_1^i\}$ where $x_1^i$ is precisely a rigid rotation of $x_0^i$ by an angle $\Delta z = z_1 - z_0$.

We can see in figure (\ref{fig:pairing-comparison}) the results of applying both our density-weighted action methodology and classical Wasserstein coupling with $c(x,y)=\tfrac12\|x-y\|^2$ (i.e., Euclidean assignment). As we can see, the former gives results much closer to the order-preserving matching that the intuition above suggests than the latter, which produce a totally different pairing, ignoring the background density and thus using chordal shortcuts across the disk. The pairing provided by the new procedure does not agree exactly with the ``ground truth'' suggested by intuition, since the paths of minimal action are not isopycnal (i.e. lines of constant density), yet it approximates it to a high degree. 

\section{Conclusions}
\label{sec:conclusion}
This article proposes an optimal transport framework where the cost of moving point $x$ to $y = T(x)$ is given by the minimal action between them, with the squared differential arclength as Lagrangian, under a metric that penalizes traversing areas of low probability. This makes both the corresponding distance between distributions and the pairing of points through the map $T$ ``natural'', in the sense that all intermediate distributions $p_t(w), \ 0 < t < 1$ between the source $p_0$ and target $p_1$ are likely to correspond to realizable scenarios. By contrast, default pairwise costs such as the squared Euclidean distance, are more susceptible to yield unnatural pairings and measures of dissimilarity among  distributions, as demonstrated through simple examples in section \cref{sec:applications}. 

Since the pairwise cost $c(x, y)$ so defined typically does not have a closed expression, one needs to simultaneously optimize over the map $y=T(x)$ pushing forward $p_0$ to $p_1$ and over the paths joining each pair $x, y$. We developed a methodology for this that parameterizes the optimal paths through Chebyshev polynomials, with path-dependent coefficients that are themselves parameterized through a reproducing kernel Hilbert space.

The methodology, tested on two-dimensional synthetic examples, is accurate and robust and extends seamlessly to general metrics and, even more generally, to arbitrary Lagrangians. It has therefore a high potential applicability not only in data analysis, but also in physics and differential geometry. Working out such applications in detail lies beyond the scope of this methodological article; it will be the subject of further work. 

\appendix
\section{Calculation details for Section 3.2}

\subsection{Left endpoint: source-side features and gradients}
This subsection assembles the source-side construction and its gradients, moving from feature building to the left-endpoint update for context and continuity.

\paragraph{Quadratic features and source side construction.}
Let $X_0=\{x_{0,j}\}_{j=1}^{n_0}$ and $\{w_i(0)\}_{i=1}^{n_0}$. Set $Y=\{w_i(0)\}\cup X_0$ with $n=n_0+n_0$ and weights
\[
f(z)=
\begin{cases}
\frac{1}{n_0}, & z=w_i(0),\\
-\frac{1}{n_0}, & z=x_{0,j},
\end{cases}
\quad \text{so } \sum_{z\in Y} f(z)=0.
\]
In two dimensions use quadratic coordinates $g(y)=(y_1,y_2,y_1^2,y_1y_2,y_2^2)$ and assemble
$G\in\mathbb{R}^{n\times 5}$ with rows $g(y)$.
Compute the column means
\[
\mu_j=\frac{1}{n}\sum_{i=1}^n G_{ij},\qquad j=1,\dots,5,
\]
stack $\mu=(\mu_1,\dots,\mu_5)^\top$, and center each column by $G_c = G - 1_n\,\mu^\top$, where $1_n$ is the $n$-vector of ones. Normalize by the Frobenius norm
\[
c = \|G_c\|_F = \sqrt{\sum_{i=1}^n\sum_{j=1}^5 G_{c,ij}^2},
\qquad
\widetilde G = \frac{G_c}{c}.
\]
Compute a truncated SVD $\widetilde G\approx U_{k_{\mathrm{svd}}} S_{k_{\mathrm{svd}}} V_{k_{\mathrm{svd}}}^\top$ and define
\[
Q_{w(0)}(y)=\widetilde G_{k_{\mathrm{svd}}}\,B_{k_{\mathrm{svd}}},\qquad B_{k_{\mathrm{svd}}}=V_{k_{\mathrm{svd}}} S_{k_{\mathrm{svd}}}^{-1}\ \ \text{(constant)}.
\]
We compute the Frobenius norm once and keep it fixed. The column means are also fixed, so they have no derivatives. Therefore
\[
\frac{\partial}{\partial w_i(0)}\,[\widetilde G_{i,j}]
=\frac{1}{\|\widetilde G\|_F}\,\frac{\partial G_{i,j}}{\partial w_i(0)},
\qquad
\frac{\partial Q_{w(0)}}{\partial w_i(0)}
=\frac{1}{\|\widetilde G\|_F}\,\frac{\partial G_{i,j}}{\partial w_i(0)}\,B_{k_{\mathrm{svd}}}.
\]
For $y=(y_{i,1},y_{i,2})=w_i(0)$, the rowwise Jacobians are
\[
\frac{\partial G_{i,1}}{\partial (y_{i,1},y_{i,2})}=(1,0),\quad
\frac{\partial G_{i,2}}{\partial (y_{i,1},y_{i,2})}=(0,1),\quad
\frac{\partial G_{i,3}}{\partial (y_{i,1},y_{i,2})}=(2y_{i,1},0),
\]
\[
\frac{\partial G_{i,4}}{\partial (y_{i,1},y_{i,2})}=(y_{i,2},y_{i,1}),\quad
\frac{\partial G_{i,5}}{\partial (y_{i,1},y_{i,2})}=(0,2y_{i,2}).
\]
The boundary contribution from the action is
\[
\frac{\partial}{\partial w_i(0)}\sum_{r=1}^{M} q_r\, L\bigl(w_{x_i}(t_r),\dot w_{x_i}(t_r)\bigr)
= -\,\frac{\partial L}{\partial \dot w}\bigl(\dot w_i(0),w_i(0)\bigr)
= -\,\frac{\dot w_i(0)}{\rho\bigl(w_i(0)\bigr)}
\quad \big(L=\tfrac12\|\dot w\|^2/\rho\big).
\]
For the penalty $\|f^\top Q_{w(0)}\|^2$,
\[
\frac{\partial}{\partial w_i(0)}\bigl\|f^\top Q_{w(0)}\bigr\|^2
= 2\,(f^\top Q_{w(0)})^\top \left(f^\top\frac{\partial Q_{w(0)}}{\partial w_i(0)}\right).
\]
Combining yields
\[
\frac{\partial \mathcal{J}}{\partial w_i(0)}
= -\,\frac{\dot w_i(0)}{\rho\bigl(w_i(0)\bigr)}
+ 2\lambda_0\,(f^\top Q_{w(0)})^\top
\left[f^\top\,\frac{1}{\|\widetilde G\|_F}\,\frac{\partial G_{i,j}}{\partial w_i(0)}\,B_{k_{\mathrm{svd}}}\right].
\]
Update rule:
\[
w(0)\leftarrow w(0)-\eta\,\frac{\partial \mathcal{J}}{\partial w(0)}.
\]
The computations at the right endpoints are entirely similar.

\subsection{Practical notes, penalty weights, and parameter updates}
Finally, we record a practical remark on statistics, provide the explicit formulas used to choose penalty weights, and list the aggregate parameter updates for completeness.

\begin{remark}
For stability and simpler code paths, compute $\mu$ and $\|G_c\|_F$ once from a fixed reference set and hold them constant during training, which yields the simple chain rule forms above. If invariance to batch level affine changes is critical, recompute the statistics at each step and use the exact Jacobian through $\widetilde{G}$.
\end{remark}

\paragraph{Choice of $\lambda$.}
Let $\sigma_0:=\|f^\top Q_0\|_2$. Balancing the typical magnitudes of the cost and penalty gradients suggests
\[
\lambda_0 \approx \frac{1}{2\sigma_*}\,
\frac{\Bigl(\sum_{i=1}^{n_0} \bigl\|\tfrac{\dot w_i(0)}{\rho\bigl(w_i(0)\bigr)}\bigr\|^2\Bigr)^{1/2}}
{\Bigl(\sum_{i=1}^{n_0} \bigl\|\tfrac{\partial \sigma_0}{\partial w_i(0)}\bigr\|^2\Bigr)^{1/2}},
\]
with
\[
\frac{\partial \sigma_0}{\partial w_i(0)}
= \frac{1}{\|f^\top Q_0\|_2}\,(f^\top Q_0)^\top
\left(f^\top\frac{1}{\|\widetilde G\|_F}\frac{\partial G}{\partial w_i(0)}\,B_{k_{\mathrm{svd}}}\right),
\]
where $Q_0=\widetilde G B_{k_{\mathrm{svd}}}$ and $\widetilde G$ is formed with fixed centering and fixed Frobenius scale. For the right endpoint define $\sigma_1:=\|f^\top Q_1\|_2$. An analogous choice is
\[
\lambda_1 \approx \frac{1}{2\sigma_*}\,
\frac{\Bigl(\sum_{i=1}^{n_0} \bigl\|\tfrac{\dot w_i(1)}{\rho\bigl(w_i(1)\bigr)}\bigr\|^2\Bigr)^{1/2}}
{\Bigl(\sum_{i=1}^{n_0} \bigl\|\tfrac{\partial \sigma_1}{\partial w_i(1)}\bigr\|^2\Bigr)^{1/2}},
\]
with
\[
\frac{\partial \sigma_1}{\partial w_i(1)}
= \frac{1}{\|f^\top Q_1\|_2}\,(f^\top Q_1)^\top
\left(f^\top\frac{1}{\|\widetilde G_1\|_F}\frac{\partial G_1}{\partial w_i(1)}\,B_{1,k_{\mathrm{svd}}}\right),
\]
where $Q_1=\widetilde G_1 B_{1,k_{\mathrm{svd}}}$ and $\widetilde G_1$ uses the fixed centering and Frobenius scale on the target side.

\paragraph{Updates.}
With step size $\eta>0$, the gradient steps are
\[
\theta_{k,j}\leftarrow \theta_{k,j}-\eta \frac{\partial \mathcal{J}}{\partial \theta_{k,j}},
\qquad
w_i(0)\leftarrow w_i(0)-\eta \frac{\partial \mathcal{J}}{\partial w_i(0)},
\qquad
w_i(1)\leftarrow w_i(1)-\eta \frac{\partial \mathcal{J}}{\partial w_i(1)}.
\]
\clearpage
\printbibliography

\end{document}